\documentclass[twocolumn,showpacs,amsmath,amssymb,pre,aps,floats]{revtex4}
\usepackage{bm,url}
\usepackage{graphicx}
\usepackage{times}
\newcommand{\BoldVec}[1]{\mathchoice%
  {\mbox{\boldmath $\displaystyle     #1$}}%
  {\mbox{\boldmath $\textstyle        #1$}}%
  {\mbox{\boldmath $\scriptstyle      #1$}}%
  {\mbox{\boldmath $\scriptscriptstyle#1$}}%
}
\newcommand{\EQ}{\begin{equation}}
\newcommand{\EN}{\end{equation}}
\newcommand{\EQA}{\begin{eqnarray}}
\newcommand{\ENA}{\end{eqnarray}}

\newcommand{\Eq}[1]{Eq.~(\ref{#1})}
\newcommand{\Eqs}[2]{Eqs.~(\ref{#1}) and~(\ref{#2})}

\newcommand{\Fig}[1]{Fig.~\ref{#1}}

\newcommand{\bra}[1]{\langle #1\rangle}

\newcommand{\mean}[1]{\overline #1}

\newcommand{\meanU}{\overline{U}}

\newcommand{\meanF}{\overline{\cal F}}
\newcommand{\meanR}{\overline{\cal R}}

\newcommand{\meanUU}{\overline{\bm{U}}}

\newcommand{\meanFF}{\overline{\mbox{\boldmath ${\cal F}$}} {}}
\newcommand{\meanRR}{\overline{\mbox{\boldmath ${\cal R}$}} {}}

\newcommand{\yyy}{\hat{\mbox{\boldmath $y$}} {}}

\newcommand{\hatU}{\hat{U}}

\newcommand{\uu}{\BoldVec{u} {}}

\newcommand{\UU}{\BoldVec{U} {}}

\newcommand{\eee}{\BoldVec{e} {}}

\newcommand{\ff}{\BoldVec{f} {}}

\newcommand{\FF}{\BoldVec{F} {}}

\newcommand{\kk}{\BoldVec{k} {}}

\newcommand{\nab}{\BoldVec{\nabla} {}}

\newcommand{\SSSS}{\bm{\mathsf{S}}}
\newcommand{\MMMM}{\bm{\mathsf{M}}}

\newcommand{\DD}{{\rm D} {}}

\newcommand{\const}{{\rm const}  {}}

\def\Rey{\mbox{\rm Re}}

\def\cs{c_{\rm s}}
\def\kf{k_{\rm f}}
\def\urms{u_{\rm rms}}
\def\Urms{U_{\rm rms}}
\def\etat{\eta_{\rm t}}

\def\half{{\textstyle{1\over2}}}

\def\onethird{{\textstyle{1\over3}}}

\newcommand{\yan}[3]{, Astron. Nachr. {\bf #2}, #3 (#1).}

\newcommand{\yana}[3]{, Astron. Astrophys. {\bf #2}, #3 (#1).}

\newcommand{\yjetp}[3]{, Sov. Phys. JETP {\bf #2}, #3 (#1).}

\newcommand{\ymn}[3]{, Mon.\ Not.\ R.\ Astron.\ Soc.\ {\bf #2}, #3 (#1).}

\newcommand{\yjfm}[3]{, J. Fluid Mech. {\bf #2}, #3 (#1).}

\newcommand{\ypre}[3]{, Phys.\ Rev.\ E {\bf #2}, #3 (#1).}
\newcommand{\yprl}[3]{, Phys.\ Rev.\ Lett.\ {\bf #2}, #3 (#1).}

\newcommand{\yaj}[3]{, Astronom. J. {\bf #2}, #3 (#1).}
\newcommand{\yapj}[3]{, Astrophys. J. {\bf #2}, #3 (#1).}

\newcommand{\ypf}[3]{, Phys. Fluids {\bf #2}, #3 (#1).}
\newcommand{\yphy}[3]{, Physica {\bf #2}, #3 (#1).}
\newcommand{\ygafd}[3]{, Geophys. Astrophys. Fluid Dynam. {\bf #2}, #3 (#1).}

\newcommand{\ybook}[3]{, {\em #2}. #3 (#1).}

\newcommand{\Sh}{{\rm Sh}}
\newcommand{\nut}{\nu_{\rm t}}
\newcommand{\nuT}{\nu_{\rm T}}
\newcommand{\nutz}{\nu_{\rm t0}}
\newcommand{\etatz}{\eta_{\rm t0}}
\graphicspath{{./fig/}{./png/}}
\begin{document}

\preprint{NORDITA 2008-44}

\title{Numerical study of large-scale vorticity generation in shear-flow turbulence}
\author{Petri J. K\"apyl\"a}
\affiliation{Observatory, T\"ahtitorninm\"aki (PO Box 14), FI-00014
University of Helsinki, Finland}
\email{petri.kapyla@helsinki.fi}

\author{Dhrubaditya Mitra}
\affiliation{Astronomy unit, School of Mathematical Sciences,
Queen Mary, University of London, Mile End Road, London E1 4NS, UK}
\email{dhruba.mitra@gmail.com}

\author{Axel Brandenburg}
\affiliation{NORDITA, Roslagstullsbacken 23, SE-10691 Stockholm, Sweden}
\email{brandenb@nordita.org}

\date{}
\begin{abstract}
Simulations of stochastically forced shear-flow turbulence in a
shearing-periodic domain are used to study the spontaneous generation
of large-scale flow patterns in the direction perpendicular to the plane
of the shear.
Based on an analysis of the resulting large-scale velocity correlations
it is argued that the mechanism behind this phenomenon could be the
mean-vorticity dynamo effect pioneered by Elperin, Kleeorin, and
Rogachevskii in 2003 (Phys.\ Rev. E 68, 016311).
This effect is based on the anisotropy of the eddy viscosity tensor.
One of its components may be able to replenish cross-stream mean flows
by acting upon the streamwise component of the mean flow.
Shear, in turn, closes the loop by acting upon the cross-stream mean
flow to produce stronger streamwise mean flows.
The diagonal component of the eddy viscosity is found to be of the
order of the rms turbulent velocity divided by the wavenumber of the
energy-carrying eddies.
\end{abstract}

\maketitle

\section{Introduction}

The imperfect analogy between the induction equation and the vorticity
equation has always raised questions regarding the extent of this analogy.
While it is well-known that the averaged induction equation for the mean
magnetic field admits self-excited solutions for a turbulent flow with
helicity, analogous solutions to the averaged vorticity equation only
exist in the compressible case \cite{Moiseev83,KRK94}.
An exception is the case of flows that are driven by a non-Galilean invariant
forcing function, which can give rise to the so-called anisotropic
kinetic alpha effect \cite{Frisch87,Sulem89,Galanti91,BvR01}.
This effect produces mean flows that are helical and of Beltrami type.
Another example of mean flow generation is the $\Lambda$ effect
\cite{Rue80,Rue89}, whereby large-scale non-uniform flows can be
produced in rotating anisotropic turbulence.

In the last few years another example has emerged, where the analogy
between vorticity and induction equations is more striking.
This example applies to the case of shear-flow turbulence.
In fact, it has been argued that large-scale magnetic field generation
is possible via the shear--current effect that results from non-vanishing
off-diagonal components of the turbulent magnetic diffusivity tensor
\cite{RK03,RK04}.
This effect predicts large-scale field generation in homogeneous
shear-flow turbulence with non-helical driving, which has indeed been
seen in several simulations \cite{Yetal08a,Yetal08b,BRRK08}.
However, there is the problem that, according to the test-field method,
the sign of the relevant component of the turbulent magnetic diffusivity
tensor was found to be incompatible with that required for the
shear--current dynamo \cite{BRRK08}.
On the other hand, the analogous hydrodynamic effect has not yet
been explored in sufficient detail.
This effect may explain the generation of large-scale vorticity
in homogeneous shear-flow turbulence and
was first studied analytically in a seminal paper by 
Elperin, Kleeorin, and Rogachevskii \cite{EKR03}. 
Several recent studies discuss numerical evidence for the spontaneous
formation of mean vorticity \cite{Yetal08a,BRRK08,Yetal08b}.
In those papers the main objective is to study the generation of large-scale
magnetic fields by shear-flow turbulence, while the simultaneous
generation of mean vorticity was merely an additional (but interesting)
complication.
On the other hand, in view of the disappointing experience when trying
to verify the operation of the shear--current dynamo using the test-field
method, one should be careful in view of earlier negative results
\cite{RK06} concerning both the shear--current effect and the
mean-vorticity dynamo effect.
The aim of this paper is therefore to discuss turbulent shear flow simulations
without magnetic fields in order to demonstrate the existence of the
mean-vorticity dynamo and to analyze its connection with the eddy
viscosity tensor in more detail.

Following earlier work \cite{Yetal08a,BRRK08,Yetal08b}, periodic boundary
conditions are used in the streamwise direction and in the direction
perpendicular to the plane of the shear, while shearing-periodic
boundary conditions are used in the cross-stream direction.
This means that mass and mean momentum are conserved.
Furthermore, if a large-scale flow emerges, it will also be periodic
corresponding to a simple sine wave.
The mean vorticity is therefore also a long-wavelength sine wave.
However, although the original analysis was based on mean vorticity,
we discuss in the following mainly the mean velocity, because the
corresponding equations are simpler and more intuitive.

For a proper analysis of the hydrodynamic mean-vorticity dynamo effect
one would need to proceed analogously to the hydromagnetic case where
it was possible to determine all relevant components of the turbulent
magnetic diffusivity tensor using the test-field method.
One would then need to determine all relevant components of the
eddy viscosity tensor.
However, in the absence of a properly developed ``test-flow'' method
for hydrodynamics, we must resort to more primitive measures for
estimating components of the eddy viscosity tensor.
Using decay calculations of a large-scale velocity structure,
it was found that eddy viscosity, $\nut$, and turbulent magnetic
diffusivity, $\etat$, are approximately equal, i.e.\ $\nut\approx\etat$,
and around $(0.8...0.9)\times\urms/\kf$ \cite{YBR03}.
Here, $\kf$ is the wavenumber corresponding to the scale of the
energy-carrying eddies and $\urms$ is the rms velocity of the turbulence.
On the other hand, a more accurate determination of $\etat$
led recently to the $\etat=\etatz\equiv\urms/(3\kf)$, where
$\etatz$ is just a reference value.
In this paper we use an analogously defined reference value,
$\nutz\equiv\urms/(3\kf)$, but note that there is no
strong case for assuming that $\nut$ will be close to $\nutz$.

\section{The model}

In the present work we consider weakly compressible subsonic turbulence
in the presence of a linear shear flow,
\EQ
\meanUU^{\rm S}=(0,Sx,0),
\EN
so $x$ is the cross-stream direction, $y$ is the streamwise direction,
and $z$ is the direction perpendicular to the plane of the shear flow.
Since the effect of temperature changes is not important in this context,
we consider an isothermal equation of state.
In the following we work with the departures from this mean flow, so the
total velocity is $\meanUU^{\rm S}+\UU$, and the governing equations
for $\UU$ are then \citep{BRRK08}
\EQ
{\DD\UU\over\DD t}=-SU_x\yyy-c_{\rm s}^2\nab\ln\rho +\ff+\FF_{\rm visc},
\label{dUU}
\EN
\EQ
{\DD\ln\rho\over\DD t}=-\nab\cdot\UU,
\label{dlnrho}
\EN
where $\DD/\DD t=\partial/\partial t+(\meanUU^{\rm S}+\UU)\cdot\nab$
is the advective derivative with respect to the total velocity,
$c_{\rm s}$ is the isothermal sound speed, here considered as constant,
$\rho$ is the mass density, $\ff$ is a random forcing function,
$\FF_{\rm visc}=\rho^{-1}\nab\cdot2\rho\nu\SSSS$ is the viscous force, and
${\sf S}_{ij}={1\over2}(U_{i,j}+U_{j,i})-{1\over3}\delta_{ij}\nab\cdot\UU$
is the traceless rate of strain tensor and commas denote partial derivatives.

The forcing function is $\delta$-correlated in time and consists of
random plane waves with wavevectors $\kk$ in the interval
$4.5\leq k/k_1\le5.5$ \cite{B01}.
During each time step, $\ff$ is a single transverse (solenoidal) plane wave
proportional to $\kk\times\eee$, where the wavevector $\kk$
is taken randomly from a set of pre-defined vectors with
components that are integer multiples of $2\pi/L$ and whose moduli
are in a certain interval around an average value, $\bra{|\kk|}$,
which we denote by $k_{\rm f}$,
and $\eee$ is an arbitrary random unit vector not aligned with $\kk$.
The corresponding scale, $2\pi/k_{\rm f}$, is referred to as
the energy-carrying scale of the turbulence.
Moreover, the time dependence of $\ff$ is designed to mimic
$\delta$-correlation, which is a simple and commonly used
form of random driving \cite{B01}.

There are two important dimensionless control parameters,
the Reynolds number $\Rey$ and the shear parameter $\Sh$,
\EQ
\Rey=\urms/(\nu\kf),\quad
\Sh=S/(\urms\kf),
\EN
that quantify the intensity of turbulence and shear, respectively.
We note that the values of $\Rey$ and $\Sh$ cannot be chosen
a priori due to the strong effect that the vorticity dynamo has on
the value of $\urms$ in the saturated state. 
Thus we always refer to values of $\urms$, $\Rey$ ,and $\Sh$ that apply
to the situation where the vorticity dynamo is absent, i.e.\ early 
stages of the run or a non-shearing simulation.
The ratio of the size of the domain, $L$, to the size of the
energy-carrying scale is also an important control parameter
that we call the scale separation ratio, written here as
$\kf L/2\pi=\kf/k_1$, where $k_1=2\pi/L$ is the smallest wavenumber
that fits into the domain.

We employ the {\sc Pencil Code} \cite{PC} with sixth-order
finite differences in space and a third order time stepping scheme.
We use triply-periodic boundary conditions, except that the $x$ direction
is shearing--periodic, i.e.\
\EQ
\UU(-\half L_x,y,z,t)=\UU(\half L_x,y+L_xSt,z,t).
\EN
This condition is routinely used in numerical studies of shear flows
in Cartesian geometry \cite{WT88,HGB95}.

\section{Results}
The initial velocity is zero, but the volume forcing drives a random
flow that soon develops a turbulent cascade where the spectral energy
follows an approximate $k^{-5/3}$ inertial range between the forcing
wavenumber $\kf$ and some dissipation wavenumber
$k_{\rm d}=\bra{\SSSS^2/\nu^2}^{1/4}$.

\begin{figure*}[t!]\begin{center}
\includegraphics[width=\textwidth]{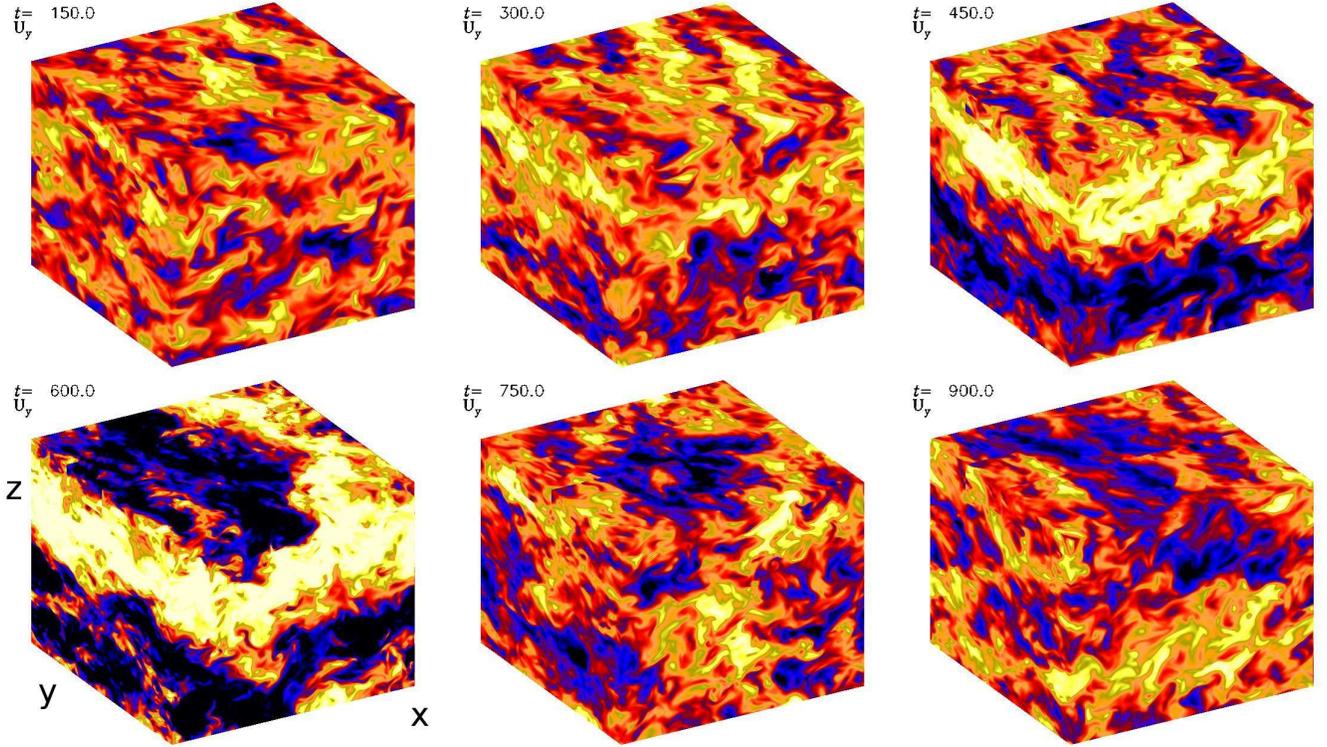}
\end{center}\caption[]{(Color online)
Representation of $U_y$ on the periphery of the computational domain
for Run~A at six different times showing the occasional generation of large
scale flow patterns with a systematic variation in the $z$ direction.
Dark (blue) shades refer to negative values of $U_y$ while light (yellow)
shades refer to positive values.
Note that at time $t\cs k_1=900$ (corresponding to $t\urms\kf\approx480$)
the orientation of the flow pattern in the $z$ direction
is reversed compared to the previous
event at $t\cs k_1=600$ (corresponding to $t\urms\kf\approx300$).
}\label{U12}\end{figure*}

In \Fig{U12} we show images of the streamwise component of $\UU$ at the
periphery of the computational domain from a run with $\Rey\approx100$ and $\Sh\approx-0.2$ (hereafter Run~A).
At early times the velocity pattern is dominated by structures whose
scale is comparable with the forcing scale, which is about one fifth
of the domain size.
However, at later times there is a tendency to produce large-scale flow patterns
with a long wavelength variation in the $z$ direction.
This flow pattern tends to be unstable and keeps disappearing
and reappearing.
This is seen also for other runs with smaller Reynolds number.

\begin{figure}[t!]\begin{center}
\includegraphics[width=\columnwidth]{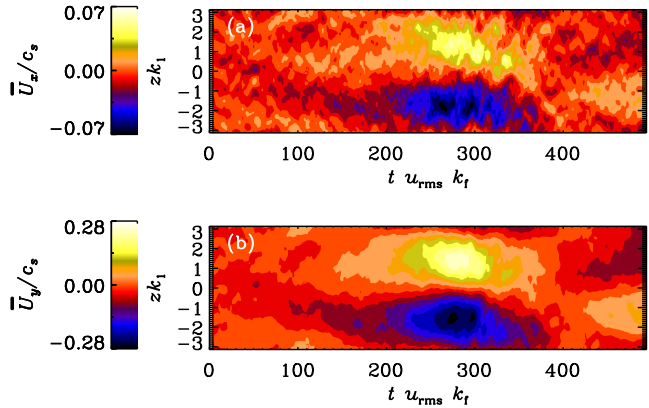}
\end{center}\caption[]{(Color online)
  $\meanU_x$ (a) and $\meanU_y$ (b) as functions of time and $z$ 
  for Run~A.}
\label{st_512a}
\end{figure}

\begin{figure}[t!]\begin{center}
\includegraphics[width=\columnwidth]{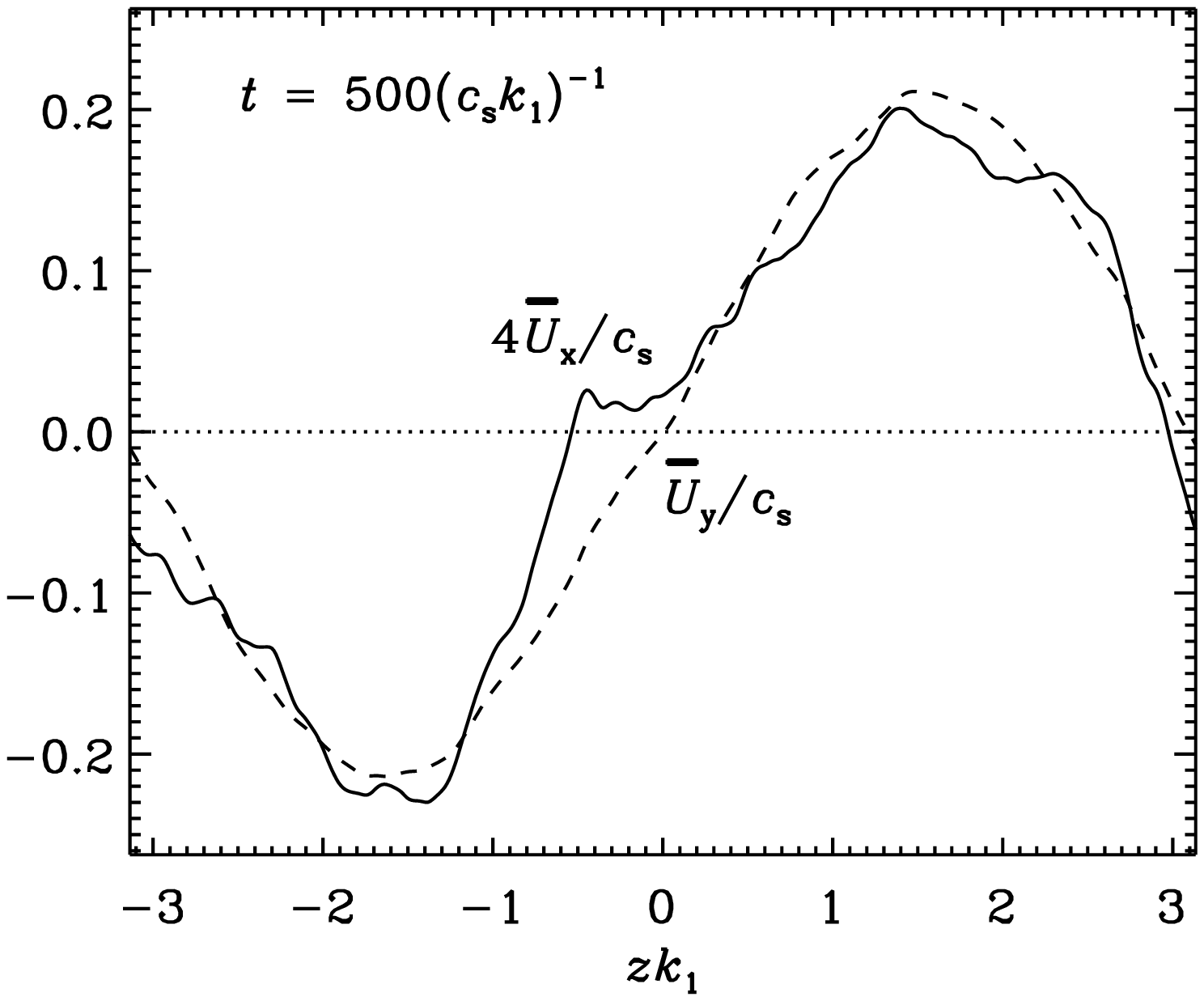}
\end{center}\caption[]{Four times $\meanU_x$ (solid line) and
  $\meanU_y$ (dashed) from Run~A at $t=500(\cs k_1)^{-1}$.}
\label{puu}
\end{figure}

Given the systematic variation in the $z$ direction, it is useful
to consider averages over the $x$ and $y$ directions, denoted in the
following by overbars.
So, $\meanUU=\meanUU(z,t)$ depends only on $z$ and $t$.
Figure \ref{st_512a} shows $\meanU_x$ and $\meanU_y$ as functions of
time and $z$.
In Fig.~\ref{puu} we plot the $z$ dependence of $\meanU_x$ and $\meanU_y$
at a time near the maximum vorticity.
Note that the amplitude of $\meanU_y$ is about 4 times as big as that of
$\meanU_x$, and that the two fields are essentially in phase.
The fact that $\meanU_x$ and $\meanU_y$ are in phase is an immediate
consequence of the fact that $S<0$, and that there is a minus sign
in front of $S$ in \Eq{dUU}.

\begin{figure}[t!]\begin{center}
\includegraphics[width=\columnwidth]{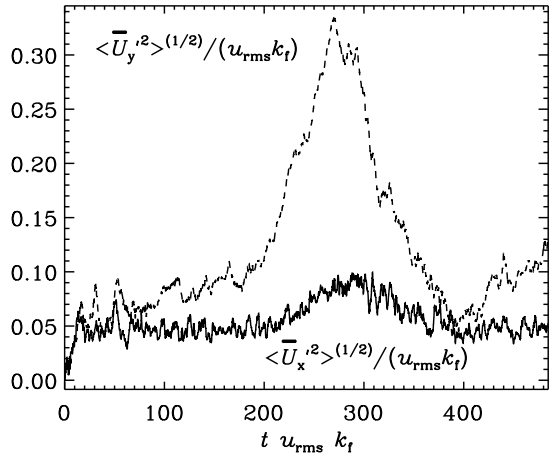}
\end{center}\caption[]{
Root mean square values of $\meanU_x'$ (solid line) and $\meanU_y'$ (dashed line) for Run~A.
Note the maxima at $t\urms\kf\approx280$ and $t\urms\kf\approx260$,
respectively.
}\label{pwrms}\end{figure}

In Run~A with $512^3$ meshpoints
there is one particularly pronounced
event during the time interval $200<t\urms\kf<400$, where $\meanU_y'$ reaches
an extremum at $t\urms\kf\approx280$, followed by an extremum of $\meanU_x'$ a bit
later at $t\urms\kf\approx300$; see \Fig{pwrms} for their root mean square values.
Here, derivatives with respect to $z$ are denoted by a prime.

The occasional extrema in the components of $\meanUU$ and its derivatives
are accompanied by strong enhancements in the rms value of the total
velocity, $\Urms$, which includes the mean flow as well.
This fact has been of some significance in previous studies of
hydromagnetic dynamo action from turbulent shear flows
\cite{Yetal08a,Yetal08b,KB08}, because,
depending on the value of the sound speed, this can lead to numerical
difficulties if the Mach number exceeds unity during these strong
enhancements of $\Urms$.
These difficulties are here avoided by choosing a smaller shear
parameter $\Sh$, regulated by the input parameter $S$.

\begin{figure}[t!]\begin{center}
\includegraphics[width=\columnwidth]{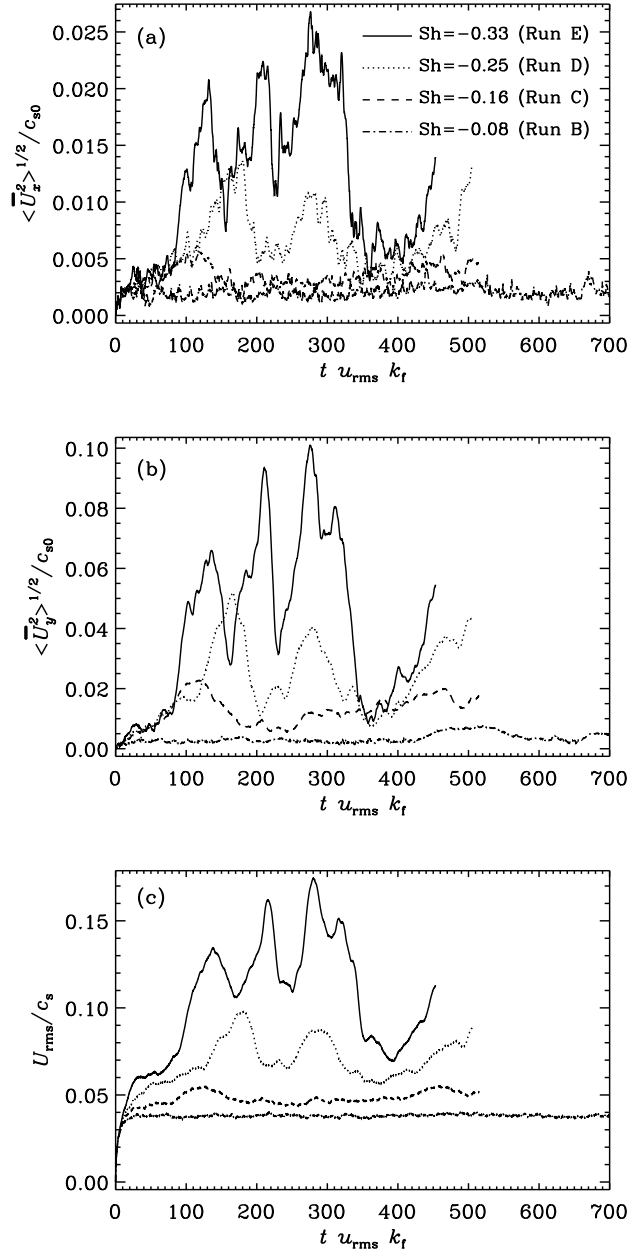}
\end{center}\caption[]{Root mean square values of $\meanU_x$ (a) and
  $\meanU_y$ (b), and $\Urms$ (c) for Runs~B to E with different shear
  as indicated in the legend in panel (a). The Reynolds number based
  on the $\urms$ from a non-shearing run is $\approx24$.}
\label{pUrms}
\end{figure}

\begin{figure}[t!]\begin{center}
\includegraphics[width=\columnwidth]{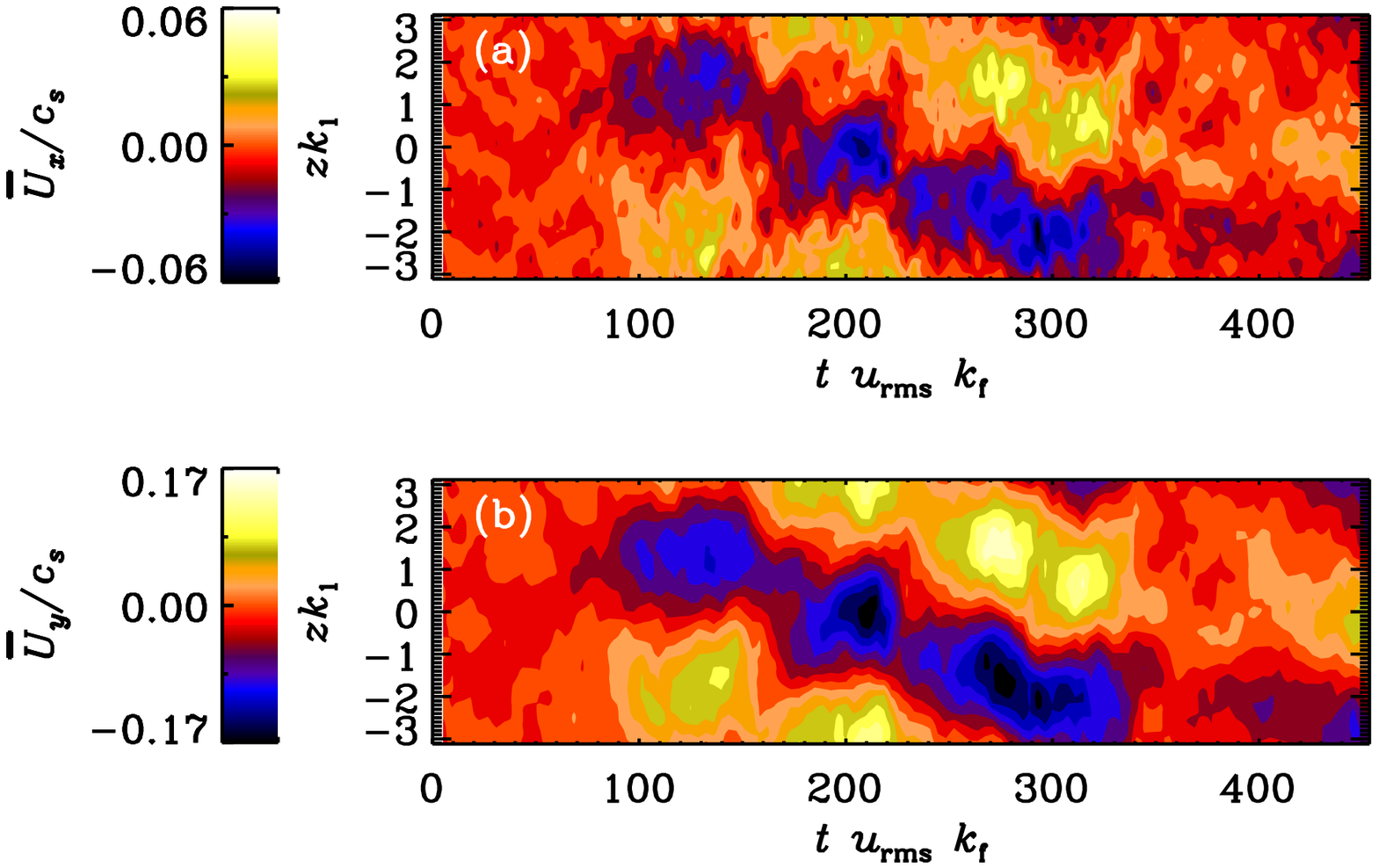}
\end{center}\caption[]{(Color online)
  $\meanU_x$ (a) and $\meanU_y$ (b) as functions of time 
  and $z$ for Run~E with $\Sh\approx-0.33$.}
\label{st_128d0}
\end{figure}

\begin{figure}[t!]\begin{center}
\includegraphics[width=\columnwidth]{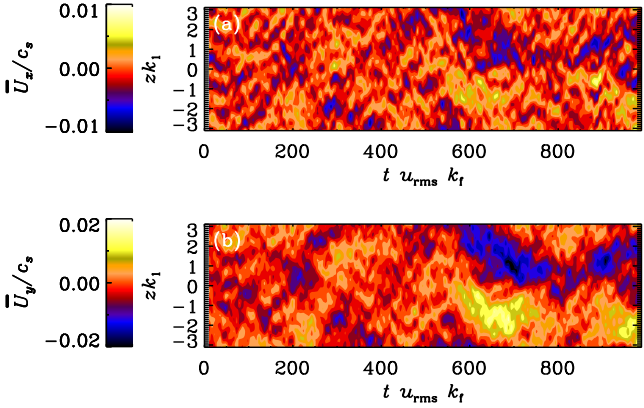}
\end{center}\caption[]{(Color online)
  $\meanU_x$ (a) and $\meanU_y$ (b) as functions of time and $z$ 
  for Run~B with $\Sh\approx-0.08$.}
\label{st_128d7}
\end{figure}

The effect of increasing $S$ is demonstrated in Fig.~\ref{pUrms},
which shows the rms values of the large-scale velocities for four runs
where $S$ is varied while the other parameters are kept constant. The
amount of shear is here quantified by the value of $\Sh$,
which is based on the $\urms$ value
from a run without shear and thus effectively quantifies the strength of the
random forcing.
These runs are denoted by the letters B to E, with the strength of the shear
increasing from $\Sh=-0.08$ in Run~B to $\Sh=-0.33$ in Run~E.
The bottom panel of Fig.~\ref{pUrms} shows that $\Urms$ increases
almost in proportion to the shear for $-\Sh>0.25$. The flow in the
large $\Sh$ runs is also highly fluctuating during periods of
vigorous vorticity generation, see Fig.~\ref{st_128d0} for a
space-time diagram of the large-scale velocities from Run~E.
Even in the lowest shear run (Run~B with $\Sh\approx-0.08$),
which is very similar to the non-shearing case during most of its evolution,
a weak
large-scale pattern is discernible at times, see times after $t\urms k_{\rm f}>600$ in Fig.~\ref{st_128d7}.

\section{Interpretation}

In order to shed some light on the mechanism responsible for the
generation of large-scale vorticity, we consider mean-field equations
\cite{EKR03,EGKR07}.
Adopting averages over the $(x,y)$ plane, denoted here by an overbar,
we have
\EQ
{\partial\meanUU\over\partial t}=-S\meanU_x\yyy+\meanFF+\nu\meanUU'',
\label{dmeanUU}
\EN
where $\meanFF=-\overline{\uu\cdot\nab\uu}$ is a term that results from the
nonli\-ne\-ari\-ty of the Navier-Stokes equations,
and primes denote a $z$ derivative.
Note that we have assumed solenoidality, i.e.\
$\nab\cdot\meanUU=\meanU_{3,3}=0$, so $\meanU_3=\const=0$
by a suitable choice of the initial condition.
Thus, only the $x$ and $y$ components of $\meanUU$ are non-vanishing.
Therefore, $\meanUU\cdot\nab\meanUU=0$.
Furthermore, the pressure gradient term does not enter in \Eq{dmeanUU}, because
any horizontally averaged gradient term can only have a $z$ component.
Using mean field theory \cite{EKR03}, $\meanFF$ can be expressed in terms
of derivatives of the mean flow.
In the present case of one-dimensional mean fields this relationship
reduces to
\EQ
\meanF_i=\nu_{ij}\meanU''_j,
\label{meanFvsmeand2Udz2}
\EN
where $\nu_{ij}$ is the eddy viscosity tensor.
We also assume incompressibility of the
small-scale velocity field, $\nab\cdot\uu=0$, which is a good
approximation for small Mach numbers, and recall that horizontal averages
depend only on $z$, i.e.\ the $j=3$ coordinate.
Therefore we have
\EQ
\meanF_i=-\nabla_3\overline{u_i u_3}=-\meanR_i'.
\EN
Here we have denoted the two relevant components of the Reynolds stress tensor
by $\meanR_i\equiv\overline{u_iu_3}$, where $i=1,2$ refer to the $x$ and $y$
directions and $u_3$ is the $z$ component of the velocity fluctuation.
Integrating \Eq{meanFvsmeand2Udz2} over $z$, we have
\EQ
\meanR_i+\nu_{ij}\meanU'_j\equiv\const.
\EN
Given that $\meanRR$ and $\meanUU$ can be obtained from the simulations,
we can then find all four components of $\nu_{ij}$ by
considering moment equations of the form
\EQ
\bra{\meanR_i\meanU'_k}+\nu_{ij}{\sf M}_{jk}=0,
\EN
where we have introduced the correlation matrix
${\sf M}_{jk}=\bra{\meanU'_j\meanU'_k}$.
We have also assumed that $\nu_{ij}$ is independent of $z$, and that, owing
to periodic boundary conditions, the mean flow and its $z$ derivatives have
zero volume average, i.e.\ $\bra{\meanU_i'}=0$ for any $i$.
The components of $\nu_{ij}$ can then be written as
\EQ
\left(\begin{array}{l}\nu_{i1}\\ \nu_{i2}\end{array}\right)=-\MMMM^{-1}
\left(\begin{array}{l}\bra{\meanR_i\meanU'_1}\\ \bra{\meanR_i\meanU'_2}\end{array}\right),
\label{nuij}
\EN
where $i=1$ or $2$.

\begin{figure}[t!]\begin{center}
\includegraphics[width=\columnwidth]{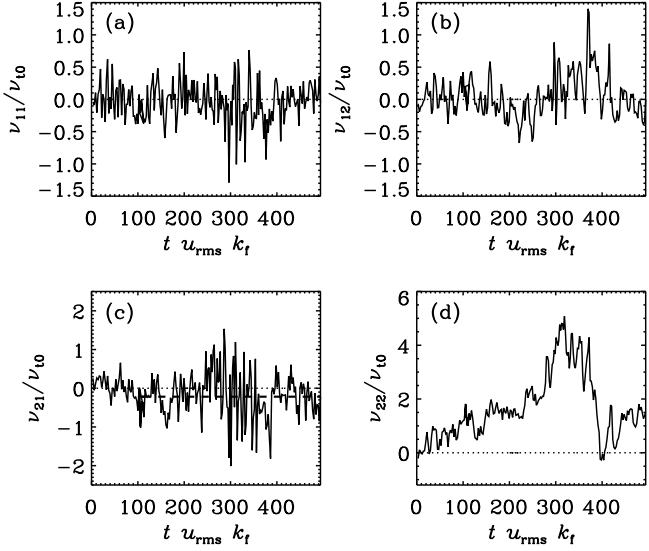}
\end{center}\caption[]{Components of the eddy viscosity tensor as
obtained from Eq.~(\ref{nuij}) normalized by $\nutz=\onethird\urms\kf^{-1}$
for Run~A.
Note that the average value of $\nu_{21}$ is negative
(see the dashed line in the third panel for $t\urms\kf>100$).
}\label{pcoefs}\end{figure}

It turns out that the components of the correlations
$\bra{\meanR_1\meanU'_j}$ are small compared with those of ${\sf M}_{jk}$.
This makes the evaluation of the components of $\nu_{1j}$ using \Eq{nuij}
ill-behaved (see Fig.~\ref{pcoefs}).
This procedure does, however, yield reasonable results for the
$\nu_{2i}$ components: $\nu_{21}$ is highly
fluctuating, but with an average of the order of roughly half of
the reference value
$\nutz\equiv\onethird\urms\kf^{-1}$, whereas $\nu_{22}$ is positive and
between one and two times $\nutz$ in the quiescent phases of the
simulation and peaking at roughly $5\nutz$ when the vorticity peaks.
The definition of $\nutz$ is analogous to a corresponding reference
value for the magnetic diffusivity \cite{Sur_etal08}, but it is not
clear that $\nut$ should be exactly equal to $\nutz$ in any limit.
Instead, according to the first order smoothing approximation,
$\nut=0.4\nutz$ \cite{YBR03}, and hence the magnetic Prandtl number was
expected to be 0.4.

\begin{figure}[t!]\begin{center}
\includegraphics[width=\columnwidth]{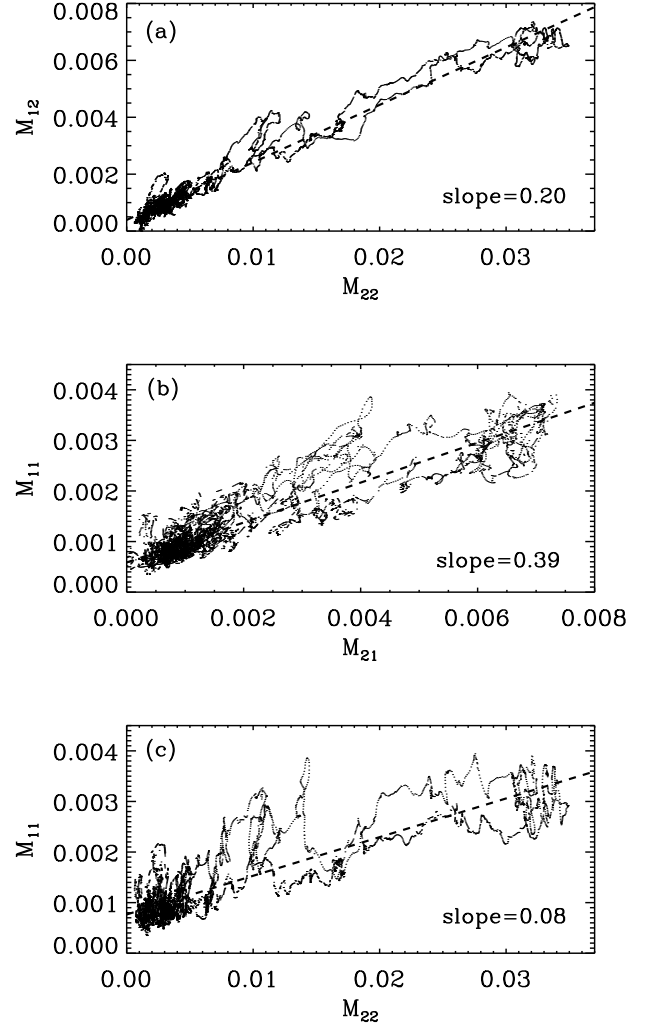}
\end{center}\caption[]{Scatter plots of ${\sf M}_{12}/{\sf M}_{22}$
  (a), ${\sf M}_{11}/{\sf M}_{21}$ (b), ${\sf M}_{11}/{\sf
    M}_{22}$ (c) for Run~A.}\label{pscatter}
\end{figure}

If both components of $\bra{\meanR_1\meanU'_k}$ for $k=1$ and 2 were
exactly zero, we could calculate $\nu_{12}/\nu_{11}$ in terms of the ratios
\EQ
{\nu_{12}\over\nu_{11}}
=-{{\sf M}_{1k}\over{\sf M}_{2k}}
\label{nu12nu11}
\EN
for $k=1$ and 2.
Yet another possibility is to take the geometric mean of the two
expressions, so
\EQ
{\nu_{12}\over\nu_{11}}\approx
-\left({{\sf M}_{11}\over{\sf M}_{21}}
\,{{\sf M}_{12}\over{\sf M}_{22}}\right)^{1/2}
\equiv-\left({{\sf M}_{11}\over{\sf M}_{22}}\right)^{1/2},
\label{m11m22}
\EN
where we have used the fact that ${\sf M}_{21}={\sf M}_{12}$.
The results shown in \Fig{pscatter} indicate that the two ratios in
Eq.~(\ref{nu12nu11}) give consistently negative
values, although their moduli are different.
Assuming that $\nu_{11}$ is positive, which is reasonable, this result
suggests that a negative $\nu_{12}$ is present in the system with a
modulus that is between 0.2 and 0.4 times the $\nu_{11}$ component.

\begin{figure}[t!]\begin{center}
\includegraphics[width=\columnwidth]{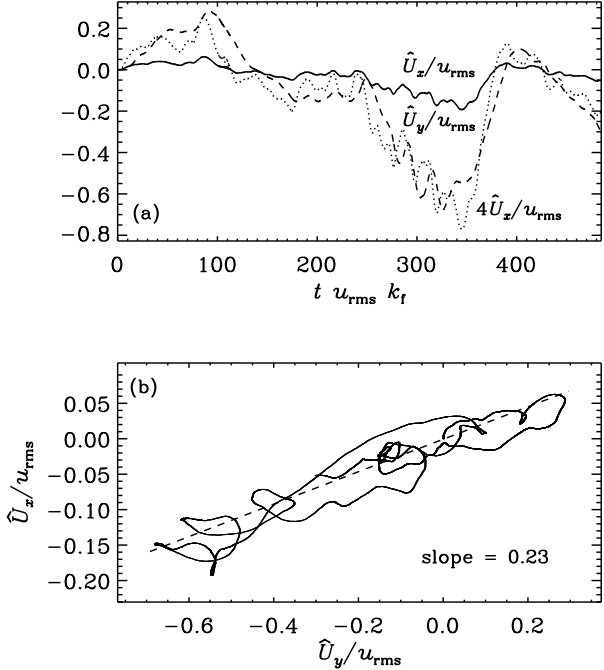}
\end{center}\caption[]{(a): $\hat{U}_x$ (solid line),
  $4\,\hat{U}_x$ (dotted) and $\hat{U}_y$ (dashed) as functions of time for
  Run~A. (b): scatter
  plot of $\hat{U}_x$ versus $\hat{U}_y$ for the same run. The dashed
  line shows a linear fit to the data.
}\label{pfamp}\end{figure}

Finally, a completely different approach for obtaining estimates
between the components of $\nu_{ij}$ is to use the resulting mean-field
equations, \Eq{dmeanUU}, and apply them to a hypothetical steady state.
These equations are linear, which is a consequence of assuming the
components of $\nu_{ij}$ to be constant.
In that case we can Fourier transform and obtain the two equations
\EQ
(\nu+\nu_{11})k^2\hatU_1+\nu_{12}k^2\hatU_2=0,
\label{F1}
\EN
\EQ
(S+\nu_{21}k^2)\hatU_1+(\nu+\nu_{22})k^2\hatU_2=0,
\label{F2}
\EN
where $\hatU_1$ and $\hatU_2$ are the Fourier amplitudes of the
$x$ and $y$ components of the mean flow.
Since these equations are linear, they cannot describe nonlinear
saturation of a mean-field vorticity dynamo instability.
However, it is plausible that the assumption of constancy
of the components $\nu_{ij}$
breaks down when the resulting mean vorticity has become large enough.
The resulting modifications of $\nu_{ij}$ may then
explain saturation.

Equations (\ref{F1}) and (\ref{F2})
show that a necessary condition for the mean-vorticity
dynamo to be excited is that the product $\nu_{12}S$ is positive.
This is indeed the case; in our case both $\nu_{12}$ and $S$ are negative.
A sufficient condition for the mean-vorticity dynamo to be excited is that
the parameter
\EQ
D\equiv\left[\nu_{12}(S/k^2+\nu_{21})+\epsilon^2\right]/\nuT^2\ge1,
\label{Ddef}
\EN
where $\nuT=\nu+\nut$ with
\EQ
\nut=\half(\nu_{11}+\nu_{22}),\quad
\epsilon=\half(\nu_{11}-\nu_{22}).
\EN
The parameter $D$ plays the role of a mean-vorticity dynamo number.
The assumption of a steady state in \Eqs{F1}{F2} implies that $D=1$.
Note that \Eqs{F1}{F2} yield
\EQ
{\nu_{12}\over\nu+\nu_{11}}=-{\hatU_1\over\hatU_2}
={\nu+\nu_{22}\over S/k^2+\nu_{21}}.
\label{nu12nu11b}
\EN
This allows us to calculate $\nu+\nu_{22}$ in terms of $\urms/\kf$,
provided $\nu_{21}$ is negligible or known:
\EQ
{\nu+\nu_{22}\over\urms/\kf}=-{\hatU_1\over\hatU_2}
\left[\Sh\left({\kf\over k_1}\right)^2+{\nu_{21}\kf\over\urms}\right].
\EN
The amplitudes $\hatU_1$ and $\hatU_2$ for Run~A are shown in Fig.~\ref{pfamp}.
Putting in numbers, $\hatU_1/\hatU_2=0.23$, $\kf/k_1=5$, we obtain
\EQ
{\nu+\nu_{22}\over\urms/\kf}=1.15-0.23{\nu_{21}\kf\over\urms},
\label{nu22}
\EN
so the uncertainty in $\nu_{21}$ enters only weakly.
Note, however, that $\nu_{22}$ is more than three times larger than
the original estimate of $\nutz$.

\section{Eddy viscosity from the imposed shear}

\begin{figure}[t!]\begin{center}
\includegraphics[width=\columnwidth]{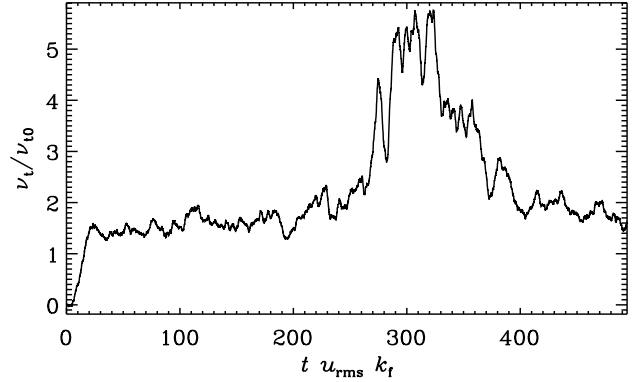}
\end{center}\caption[]{Turbulent viscosity for Run~A, as obtained from the
Reynolds stress component $\mean{R}_{xy}$, divided by the estimate $\nutz$.
}\label{pnut}\end{figure}

We have so far only looked at the components of the Reynolds stress
tensor that enter the horizontally averaged equations.
However, there is at least one other component that does not enter
\Eq{dmeanUU}, but that can also be used to determine the eddy viscosity (see, e.g.\ \cite{KB07}).
This component is not driven by the derivatives of $\meanUU$, but by
the imposed shear flow, $\nabla_x\meanU_y^S$, itself.
Indeed, one expects that this imposed shear leads to an $xy$ stress
\EQ
\overline{u_xu_y}=-\nut\left(\nabla_x\meanU_y^S+\nabla_y\meanU_x^S\right)
=-\nut S.
\label{uxuy}
\EN
This is indeed the case; see \Fig{pnut}.
It turns out that the $\nut$ determined in this way is rather similar
to the value of $\nu_{22}$ estimated from \Eq{nu22}.
Again, there is no good reason that these values are the same,
because the eddy viscosity obtained from \Eq{uxuy} belongs to a
different component of the full rank-4 eddy viscosity tensor and
is not part of the rank-2 tensor considered above.

\section{Conclusions}

The present work has demonstrated quite clearly that in non-helical
shear-flow turbulence a large-scale flow pattern emerges spontaneously.
In the present case, where in the absence of shear the turbulence
saturates at a Mach number of order 0.01, the large-scale flow becomes
exceedingly strong and saturates at a Mach number of 0.1--0.2.
This behavior is seen both at small and at the largest Reynolds numbers
considered here ($\Rey=100$, based on the inverse forcing wavenumber).

The flow pattern can be particularly well pronounced at certain times
and shows a long wavelength variation in the direction perpendicular
to the plane of the shear flow (here the $z$ direction).
For negative shear, the $x$ and $y$ components of the shear flow are in
phase in a way that is compatible with an interpretation in terms of a
large-scale vorticity dynamo, as explored first by Elperin, Kleeorin,
and Rogachevskii \cite{EKR03}.
This means that the large-scale flow is driven by an anisotropic
eddy viscosity tensor.
Particularly important is its $xy$ component, $\nu_{xy}$, which
describes the production of a cross-stream large-scale flow component,
$\meanU_x(z,t)$, from a $z$-variation of the streamwise large-scale flow,
$\meanU_y(z,t)$.
The mean-vorticity ``dynamo cycle'' is completed by a suitable action of the
shear itself, which produces a streamwise large-scale flow component,
$\meanU_y$, from the cross-stream component, $\meanU_x$, by the term
$-\meanUU\cdot\nab\meanUU^S$.

The mean-vorticity dynamo cycle can only work if the sign of $\nu_{xy}$
is the same as that of the shear, $\nabla_x\meanU_y^S$.
The present investigations suggest that this is indeed the case.
However, it is desirable to verify the sign of $\nu_{xy}$ using
a test-flow method analogously to the test-field method used in
magnetohydrodynamics.
Some care in using the correlation method is in order, because there
are examples in magnetohydrodynamics where the correlation method
give incorrect values for some components of the magnetic diffusion
tensor, although other components were correct \cite{BS02}.
For example, when we apply a method analogous to that in \Eq{nu12nu11b}
to the magnetic field of a simulation of shear flow turbulence (see, e.g.,
Figs.~7 or 8 of Ref.~\cite{BRRK08}), the components of the magnetic field
scatter almost isotropically about the origin.
This is compatible with an interpretation in terms of an incoherent
alpha--shear effect \cite{BRRK08,VB97}.
On the other hand, there is still a weak correlation with a negative slope.
This would suggest that the shear--current dynamo might also be at work,
even though the test-field method indicates that this should not be
the case.

Clearly, the reality of the large-scale flow found in simulations is more
complicated than what is suggested by the simple mean-vorticity dynamo problem.
Firstly, in contrast to the magnetic dynamo no kinematic stage can be
distinguished, i.e.\ the large-scale patterns are visible only after
they are already of dynamical importance.
Secondly, the mean flow can reverse sign
in random intervals 
which is not anticipated from the linear mean-vorticity dynamo model with
anisotropic eddy viscosity, where self-excited solutions would always
be non-oscillatory.
Another question that needs to be addressed in future work is the
saturation level of the large-scale flow, its relation to the saturation
level of the small-scale flow, and a possible dependence on the Mach
number.

For more realistic applications it will be important to get information
about the full eddy viscosity, which is a rank-4 tensor \cite{Rue89}.
In the present work, where the averages are only one-dimensional,
the eddy viscosity reduces to a rank-2 tensor.
Finally, for astrophysical applications it should be pointed out that
the gas in many shear flows is ionized and electrically conducting,
giving rise to efficient dynamo action.
The resulting mean Lorentz force from the small-scale magnetic field
modifies the eddy viscosity in a way that suppresses the mean-vorticity dynamo.
Details of this need to be investigated further.
Another effect that can suppress the mean-vorticity dynamo is rotation
\cite{Yetal08b}.
This can be understood from the dispersion relation in that the
addition of rotation leads, among other terms,
to a $-4\Omega^2$ term inside the squared
brackets of \Eq{Ddef} that always suppresses the mean-vorticity dynamo.

\acknowledgements
We thank an anonymous referee for offering suggestions regarding the
break-down of the linearity of equations (\ref{F1}) and (\ref{F2}).
The computations were performed on the facilities hosted by the
Center of Scientific Computing in Espoo, Finland, who are administered
by the Finnish ministry of education.
This work was supported by the Academy of Finland grant No.\ 121431 (PJK),
the Leverhulme Trust (DM), and the Swedish Research Council (AB).

\vfill\bigskip\noindent{\it
$ $Id: paper.tex,v 1.74 2008-12-31 08:57:22 brandenb Exp $ $}

\end{document}